\documentclass[aps,pre,reprint,amsmath,amssymb]{revtex4-2}
\usepackage{txfonts}
\usepackage{algorithmic}
\usepackage{algorithm}
\usepackage{graphicx}
\usepackage{bm}
\usepackage{url}

\newcommand{\argmax}{\mathop{\rm arg~max}\limits}
\newcommand{\argmin}{\mathop{\rm arg~min}\limits}

\begin{document}

\title{Characterization of Locality in Spin States and Forced Moves for Optimizations}

\author{Yoshiki Sato$^1$, Makiko Konoshima$^2$, Hirotaka Tamura$^3$, and Jun Ohkubo$^1$}

\affiliation{$^1$Graduate School of Science and Engineering, Saitama University, Sakura, Saitama, 338--8570 Japan\\
  $^2$Fujitsu Limited, Kawasaki, Kanagawa, 211--8588 Japan\\
$^3$DXR laboratory Inc., Yokohama, Kanagawa, 223--0066 Japan}

\begin{abstract}
Ising formulations are widely utilized to solve combinatorial optimization problems, and a variety of quantum or semiconductor-based hardware has recently been made available. In combinatorial optimization problems, the existence of local minima in energy landscapes is problematic to use to seek the global minimum. We note that the aim of the optimization is not to obtain exact samplings from the Boltzmann distribution, and there is thus no need to satisfy detailed balance conditions. In light of this fact, we develop an algorithm to get out of the local minima efficiently while it does not yield the exact samplings. For this purpose, we utilize a feature that characterizes locality in the current state, which is easy to obtain with a type of specialized hardware. Furthermore, as the proposed algorithm is based on a rejection-free algorithm, the computational cost is low. In this work, after presenting the details of the proposed algorithm, we report the results of numerical experiments that demonstrate the effectiveness of the proposed feature and algorithm.

\vspace{10mm}$\,$
\end{abstract}

\maketitle

\section{Introduction}
\label{introduction}

Ising models are widely utilized in physics, and many works have investigated their practical applications. A recent hot topic in this domain is the combinatorial optimization problem, where the goal is to find an optimal state combination that minimizes the energy or cost function of the problem. The greater the number of state combinations, the exponentially longer it takes to find the global minimum; therefore, a brute-force method of calculating the energy exhaustively for all states becomes impractical. Specialized hardware has therefore been developed in recent years to solve the combinatorial optimization problem at high speed. A typical example is D-Wave Advantage by D-Wave Systems \cite{Bunyk2014,d-wave,d-wave-company}, which harnesses the principle of quantum annealing. Another example encompasses technologies based on complementary metal-oxide semiconductors (CMOS), including devices developed by Hitachi \cite{Takemoto2019,Takemoto2021}, Toshiba \cite{Goto2019a,Goto2019b,Tatsumura2019}, and Fujitsu \cite{Aramon2019,Matsubara2020} As the input format, these annealing machines typically use a quadratic form of binary variables called quadratic unconstrained binary optimization (QUBO) formulation, which is equivalent to the Ising model. Various combinatorial optimization problems have been reformulated as QUBO formulations \cite{Lucas2014}.

There are many practical applications of the QUBO formulation and specialized hardware. Applications for optimization include the radiation dose in radiotherapy \cite{Nazareth2015} and traffic flows in cities \cite{Neukart2017,Ohzeki2019}. Various machine-learning methods have also been discussed, include those that deal with clustering problems \cite{Kumar2018,Arthur2021} and black-box optimization \cite{Izawa2022}. Estimation of the structures of polymers is also possible \cite{Endo2022}. As for the various applications in quantum annealing, readers are directed to the examples in Yarkoni et al.'s review paper \cite{Yarkoni2022}.

In combinatorial optimizations, local minima in energy landscapes can be problematic. Specifically, the state variable may fall into local solutions and become trapped there, which impedes the search for the optimal solution. There are many heuristic algorithms for optimization problems, which are implemented as software \cite{Korte_book}. Since the heuristics vary from problem to problem, the development of specialized hardware has been explored, but the local minima are still problematic even in specialized hardware. In addition, it is not easy to implement heuristic algorithms into hardware, especially in quantum annealers.

In this paper, we propose a new algorithm to escape the local minima, assuming future implementations into CMOS-type hardware. The key element of our algorithm is its rejection-free selection \cite{Rosenthal2021}: namely, it implements an update rule analogous to rejection-free selection in the opposite manner. Rejection-free selection has already been successfully implemented on dedicated hardware and highly parallel platforms such as GPUs \cite{Aramon2019}. Therefore, the various quantities required to perform rejection-free selection are readily available for other uses on such computing platforms, with minimal additional computational overhead. Although the change of the update rule makes exact sampling from the Boltzmann distribution impossible, the aim of the optimization is to seek the global minimum. Hence, there is no need to satisfy detailed balance conditions. By exploiting this fact, the proposed algorithm is able to get out of the local minima efficiently while it does not yield the exact samplings. In the algorithm, we utilize a feature that characterizes locality in the current state, which is an easy feature to obtain with specialized hardware. We also discuss the meaning of the proposed algorithm with a probabilistic interpretation.

Section~II of this paper reviews the annealing machine formulation and the replica exchange method. Our main proposals, namely, the characterization of locality, a method to escape from local minima, and their interpretations, are presented in Sect.~III. In Sect.~IV, we report the results of numerical experiments that demonstrate the effectiveness of the proposed algorithm. We conclude in Sect.~V with a brief summary and mention of future work.

\vspace{10mm}

\section{Backgrounds of Annealing Machines}

\subsection{Ising model and QUBO formulation}

The Ising model is a fundamental model to describe the properties of magnetic materials. Let $\sigma_i \in \{-1, 1\}$ denote a variable for the $i$-th spin. Then, the energy function of the Ising model of $N$ spins with the state vector $\bm{\sigma}$ is defined as
\begin{align}
  \label{eq:ising_model}
  E(\bm{\sigma}) = -\frac{1}{2}\sum_{i \in \mathcal{D}}\sum_{j \in \mathcal{D}} J_{ij} \sigma_i \sigma_j - \sum_{i \in \mathcal{D}} h_i \sigma_i,
\end{align}
where $\mathcal{D}$ is the set of indices of the variables, $J_{ij} \in \mathbb{R}$ corresponds to the two-body interaction between the spins $\sigma_i$ and $\sigma_j$, and $h_i \in \mathbb{R}$ is the external magnetic field on $\sigma_i$. Note that $|\mathcal{D}| = N$.

The QUBO formulation has the following cost function for the state vector $\bm{z}$:
\begin{align}
  E(\bm{z}) = -\sum_{i \in \mathcal{D}}\sum_{j \in \mathcal{D}}Q_{ij}z_iz_j,
  \label{eq:qubo}
\end{align}
where $z_i \in \{0,1\}$ is the $i$-th binary variable in $\bm{z}$, and $Q_{ij} \in \mathbb{R}$ is the strength of the interaction between the binary variables $z_i$ and $z_j$. This formulation is equivalent to the Ising model via the variable transformation with 
\begin{align}
  z_i = \frac{1 + \sigma_i}{2}.
  \label{eq:variable_change}
\end{align}
Conversion between $\{J_{ij}\}, \{h_i\}$, and $\{Q_{ij}\}$ is also possible.

From the computational viewpoint, annealing machines are domain-specific, and their role is simple: essentially, they are tasked with finding the ground state minimizing Eq.~\eqref{eq:ising_model} or \eqref{eq:qubo}. Despite the domain-specific characteristic, we can solve various combinatorial optimization problems using annealing machines. This is because the given optimization problem is transformed into an Ising or QUBO form so that the ground state coincides with the optimal solution of the combinatorial optimization problem \cite{Lucas2014}.

\subsection{Annealing mechanism}

Although quantum annealing \cite{Kadowaki1998,Farhi2001} is a well-known mechanism for annealing machines, it is difficult to implement some heuristics into specialized hardware due to its quantum nature. In the current work, we try to improve the algorithms in classical annealing schemes with an eye toward future implementation on specialized hardware. Hence, we give brief reviews on classical annealing mechanisms in the following subsections.

\subsubsection{Simulated annealing}

Simulated annealing (SA) is an algorithm based on thermal fluctuations. When the temperature is high, the thermal fluctuations are large. The algorithm searches the large area in the state spaces and settles to the ground state by gradually decreasing the temperature. Numerically, the probability $P$ of a state change is typically defined as
\begin{align}
P = \min\left[1, \exp\left(-\frac{\Delta E}{T}\right)\right],
\label{eq:probability}
\end{align}
where $T$ is the temperature and $\Delta E$ is the energy difference from the previous state to the next one. This is the conventional Metropolis rule.

We can find the ground state if the temperature decreases slowly enough \cite{Geman1984}. However, in practical cases using SA, temperatures are often lowered faster than this rate; the rate described in Ref.~\cite{Geman1984} is too slow and impractical.

\subsubsection{Replica exchange method}

The replica exchange method \cite{Swendsen1986,Hukushima1996}, also known as the parallel tempering method, is well-known for its ability to improve the sampling efficiency of Monte Carlo simulations and Markov chain Monte Carlo methods. In the replica exchange method, the temperature of each replica is determined and fixed in advance. Conventional Monte Carlo sampling is typically performed on each replica, and sometimes the state variables between randomly selected replicas are exchanged. In physics, the replica exchange method is often utilized to find ground energy states, and it is also used in some specialized hardware for optimization problems.

In the replica exchange method, each replica develops with the conventional Metropolis rule in Eq.~\eqref{eq:probability}. Since each replica has a different temperature, we set $T_{m}$ as the temperature of the $m$-th replica. Then, the probability of state change in the $m$-th replica, $P_{\mathrm{change}}^{(m)}$, is defined as
\begin{align}
P_{\mathrm{change}}^{(m)} = \min\left[1, \exp\left(-\frac{\Delta E^{(m)}}{T_{m}}\right)\right],
\label{p_change}
\end{align}
where $\Delta E^{(m)}$ is the energy difference from the previous state to the next one in the $m$-th replica.

The essential feature here is the exchange of state variables between different replicas. Although low-temperature settings are necessary when seeking stable states, there are many local minima that have difficulty escaping at low temperatures. Therefore, using the replicas with high temperatures, we can seek various configurations.

The probability $P_{\mathrm{exchange}}^{(m,l)}$ of an exchange occurring at the $m$-th and $l$-th replicas is defined as
\begin{align}
  P_{\mathrm{exchange}}^{(m,l)} = \mathrm{min}\left[1,\exp\biggl\{(E^{(m)} - E^{(l)})\biggl(\frac{1}{T_{m}} - \frac{1}{T_{l}}\biggr)\biggr\}\right],
  \label{p_exchange}
\end{align}
where $E^{(m)}$ and $E^{(l)}$ represent the energies of the $m$-th and $l$-th replicas, respectively.

In the present paper, the temperature is ordered in ascending order from lowest to highest, i.e., $T_1 < T_2 < \dots < T_M$, where $M$ is the number of replicas.  We will consider only exchanges at two adjacent replicas, and the following equation holds in Eq.~\eqref{p_exchange}:
\begin{align}
l = m + 1 \qquad (m = 1, \dots, M-1).
\end{align}

Note that the proposed algorithm discussed in Sect.~III is suitable for the replica exchange method.

\subsubsection{Comment on computational costs}

We here comment on the computational costs of calculating the energy difference, which is needed for performing the update procedures. The energy is defined by Eq.~\eqref{eq:ising_model} or \eqref{eq:qubo}, and hence the computational complexity is $\mathcal{O}(N^2)$; in other words, it is a little time-consuming.

A practical approach to evaluate the energy difference is to introduce $\Delta \sigma_{i}$ with $\Delta \sigma_{i} = -2\sigma_{i}$. Assume that the $i$-th spin is chosen as the candidate spin in the Metropolis rule. Then, the energy difference $\Delta E$ is calculated as
\begin{align}
\Delta E = -\Delta \sigma_{i}\left(\sum_{j \in \mathcal{D}} J_{ij} \sigma_j + h_i\right).
\label{delta_Ising}
\end{align}
As for the QUBO formulation, the introduction of $\Delta z_{i}$ with $\Delta z_{i} = 1 - 2z_{i}$ immediately yields the concise expression for the energy difference $\Delta E$ as
\begin{align}
\Delta E = -\Delta z_{i}\left(\sum_{j \in \mathcal{D}}Q_{ij}z_j \right).
\label{delta_QUBO}
\end{align}
Then, Eq.~\eqref{delta_Ising} or \eqref{delta_QUBO} reduces the computational complexity from $\mathcal{O}(N^2)$ to $\mathcal{O}(N)$.

\section{Proposals of Characterization of Locality and Algorithm with Forced Moves}

When a state is trapped in a local minimum, a naive algorithm is to re-initialize the state randomly and re-start the Monte Carlo method. However, we confirmed through preliminary experiments that random re-initialization does not improve the optimization result. In this section, we propose and discuss an approach that would be suitable for future hardware implementation.

\subsection{Outline of the proposal}

In this section, we propose an algorithm to enhance the escape from local minima. Although there are many candidates for such an algorithm, it is preferable that it have the following properties:
\begin{itemize}
\item Low computational cost
\item Interpretability
\end{itemize}

\begin{figure}[b]
  \centering
  \includegraphics[keepaspectratio,width=80mm]{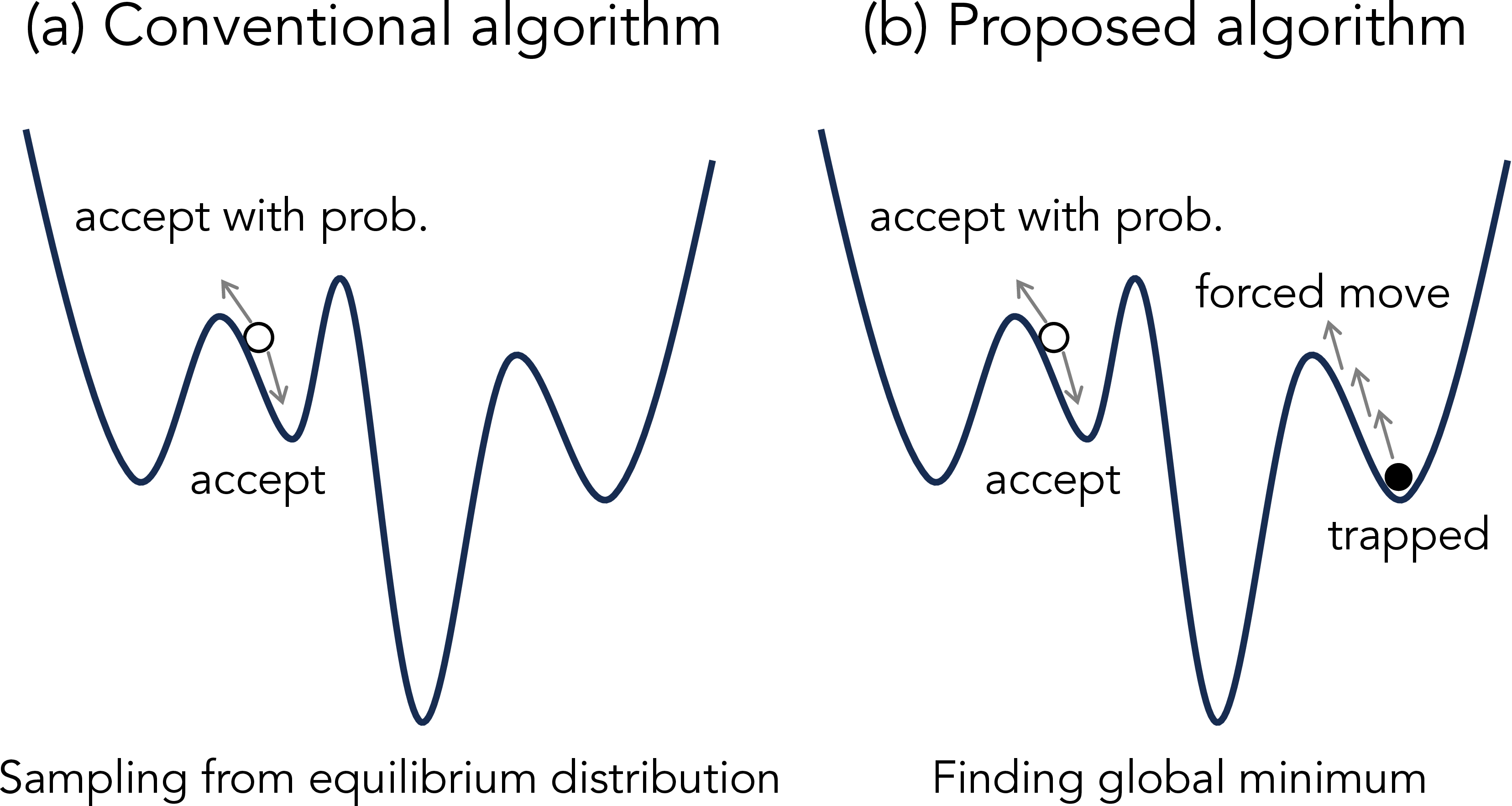}
\caption{Conceptual explanations of (a) the conventional Monte Carlo sampling and (b) our proposed algorithm with forced moves. We judge whether the state is in local minima or not. If it falls into the local minima, forced moves are applied.}
  \label{fig:concept}
\end{figure}

Assuming future implementation in hardware, we propose a characterization of locality and an algorithm so that the above two properties are satisfied. Figure~\ref{fig:concept}(a) and (b) depict conceptual explanations of the conventional and proposed algorithms, respectively. As shown in (b), the proposed algorithm has forced moves to escape the local minima. Our approach utilizes a characterization of locality, which can judge whether the state is trapped into local minima nor not. As discussed in Sect.~I, the forced moves violate the local detailed balance condition. However, our aim is to solve the optimization problems, not to sample equilibrium distributions. 

\subsection{Characterization of locality}

We first characterize the locality, as its qualification is necessary to apply the forced moves in the algorithm. Here, we use the following quantity as the characteristics of locality:
\begin{align}
P_{\mathrm{escape}} = \frac{1}{N}\sum_{i=1}^{N} \mathrm{min}\biggl(1,\exp\biggl\{{-\frac{\Delta E_{i}}{T}}\biggr\}\biggr),
\label{escape_prob}
\end{align}
where $\Delta E_{i}$ represents the energy difference of the $i$-th spin flip. 

As discussed in Sect.~II.B.3, each energy difference $\Delta E_i$ is evaluated in $\mathcal{O}(N)$. Therefore, the total computational cost of calculating $P_{\mathrm{escape}}$ is $\mathcal{O}(N^2)$. Although this cost is slightly high, it can be reduced by using specialized hardware. For example, the calculation of Eq.~\eqref{delta_QUBO} can be performed faster in Fujitsu Digital Annealer \cite{Aramon2019,Matsubara2020} due to hardware parallelization, which would make it possible to evaluate the quantity $P_{\mathrm{escape}}$ in a sufficiently realistic time.

The interpretability of the quantity $P_{\mathrm{escape}}$ is obvious: it corresponds to the average escape (or transition) probability from the current configuration. A smaller $P_{\mathrm{escape}}$ means that it is more difficult to escape from the current state. When $P_{\mathrm{escape}}$ is large, we can judge that the locality at the current state is low. 

For the above reasons, it makes sense to utilize the escape probability as the characteristic of the locality.

\subsection{Proposed algorithm}

Next, we propose an algorithm using the characteristic, i.e., the average escape probability $P_{\mathrm{escape}}$. The basic idea is simple: when the state is trapped in a local minimum, we apply forced moves until the average escape probability is large enough.

\begin{algorithm}[H]
\caption{Processing for forced moves}
\label{alg1}
\begin{algorithmic}
  \STATE (Perform conventional update procedures)
  \IF{The state is trapped in a local minimum}
    \WHILE{$P_{\mathrm{escape}} \le \alpha$}
    \STATE Select a spin variable $j$ via \\$\quad j = \argmax_i \left[\max\left(0, \Delta E_i \right) + T \log \left(-\log(s_i) \right) \right]$
    \STATE Flip $j$-th spin
    \ENDWHILE
  \ENDIF
\end{algorithmic}
\end{algorithm}


First, we apply the conventional Metropolis update rules, which sometimes reject several consecutive spin flips. When the rejections continue to some extent, we judge that the current state is trapped in a local minimum. Although there are other possible criteria to judge this state, we utilize the simple criterion since this judgment is not essential to the proposals.

Second, we define the forced moves to escape from the local minima. In the forced moves, a spin variable to be flipped is selected according to the following rule:
\begin{align}
j = \argmax_i \left[\max\left(0, \Delta E_i \right) + T \log \left(-\log(s_i) \right) \right],
\label{select_rule}
\end{align}
where $\Delta E_i$ is the energy difference for the $i$-th spin flip, and $s_i$ is a uniformly distributed random number with $0 < s_i < 1$. The random numbers $s_1, \dots, s_N$ are independent of each other. This rule is analogous to the rejection-free selection rule \cite{Rosenthal2021}. The rejection-free selection rule uses $\argmin$ instead of $\argmax$ in Eq.~\eqref{select_rule}, which enhances the exploitation via the conventional Metropolis rule. Here, we aim to escape from the local minima, so we apply the rejection-free selection rule in an opposite manner. The reason for this choice is tied to the future implementation of hardware: specifically, the rejection-free selection rule is suitable for implementation in certain specialized hardware \cite{Aramon2019,Matsubara2020}, so we expect the rule in Eq.~\eqref{select_rule} to be easy to implement in the future. Then, the reduction of computational costs in hardware would be possible. We will discuss the meaning of the rule in more detail later.

Third, we repeat the forced moves until we reach different configuration spaces near the trapped minimum. Then, the averaged escape probability in Eq.~\eqref{escape_prob} is used to judge whether the updated spin configuration is far enough away from the trapped minimum. If the criterion is not satisfied, we successively select the next spin variables by Eq.~\eqref{select_rule}, and the forced moves are implemented.

Algorithm~1 summarizes the forced moves. This algorithm has a control parameter $\alpha$ that determines the distance from the local minimum. If the criterion $P_{\mathrm{escape}} > \alpha$ is not satisfied, the forced moves are applied successively. After the criterion is satisfied, the conventional Metropolis update rule is implemented again.

Note that the proposed method is applicable to SA, as the forced moves enhance the escape from the local minima. However, the temperature gradually decreases in SA, and at low temperature, the spin configuration is more likely to be pulled back to the local minimum immediately after finishing the forced moves. Hence, in the present paper, we apply the proposed method to the replica exchange method. 

In the replica exchange method, the judgments of the local trapping and forced moves are applied to each replica independently. As discussed above, we here utilize a simple judgment rule for the local trapping: when there are successive rejections in the Metropolis update rule, the current state is in a local minimum. In the numerical experiments discussed later, $20$ successive rejections are utilized as the judgment criterion. Note that we exchange both the spin configurations and the number of rejections in the replica exchange.

\subsection{Interpretation of the rule in forced moves}

Here, we discuss the interpretation of the selection rule in Eq.~\eqref{select_rule}. As stated above, the rule is analogous to the rejection-free selection rule, in which $\argmin$ is used instead of $\argmax$. The original rejection-free selection rule explicitly selects the $j$-th node in proportion to the probability in the Metropolis rule (see the discussion in the Appendix of Ref.~\cite{Yin2023} for more details). 

As a first step in the discussion, we define the following quantity:
\begin{align}
A_i = \min\left[ 1, \exp\left(- \frac{\Delta E_i}{T} \right)\right],
\end{align}
where $\Delta E_i$ is the energy difference for the $i$-th spin flip. As easily seen, it corresponds to the conventional Metropolis rule. 

Let $\{s_i\}$ be independent uniformly distributed random numbers with $0 < s_i < 1$. We here focus on the quantity $s_i^{-1/A_i}$. If $s_i$ is fixed, $s_i^{-1/A_i}$ increases as $A_i$ decreases. A smaller value of $A_i$ means that the transition via the $i$-th spin flip is less likely to occur. Hence, we can use the quantity $s_i^{-1/A_i}$ to select a spin with a small probability of transitions, which leads to a spin that is difficult to get out of the local minimum. Of course, the random number $s_i$ enables us to introduce randomness to the selection rule.

According to the above discussion, we utilize the following selection rule:
\begin{align}
j = \argmax_i \left[ s_i^{-1/A_i}\right].
\end{align}
Note that the logarithmic function is a monotonically increasing function. Hence, the right-hand side is rewritten as
\begin{align}
\argmax_i \left[ s_i^{-1/A_i}\right] 
&= \argmax_i \left[ \log\left( s_i^{-1/A_i} \right) \right] \nonumber \\
&= \argmax_i \left[ \frac{- \log s_i}{A_i} \right] \nonumber \\
&= \argmax_i \left[ \log\left( \frac{- \log s_i}{A_i} \right) \right] \nonumber \\
&= \argmax_i \left[ \log\left( - \log s_i\right) - \log A_i \right] \nonumber \\
&= \argmax_i \left[ \log\left( - \log s_i\right) - \min\left[0, \left(- \frac{\Delta E_i}{T} \right)\right] \right] \nonumber \\
&= \argmax_i \left[ T \log\left( - \log s_i\right) + \max\left(0, \Delta E_i \right) \right],
\end{align}
which is consistent with the selection rule in Eq.~\eqref{select_rule}.

On the other hand, we consider the probability $P_j$ that the rule in Eq.~\eqref{select_rule} selects the spin number $j$. This probability is explicitly evaluated as follows:
\begin{align}
P_j &= \mathrm{Prob}\left( \argmax_i \left[ s_i^{-1/A_i}\right]  = j\right) \nonumber \\
&= \mathrm{Prob}\left( \bigcap_{i\neq j}  \left[ s_i^{-1/A_i}  \leq s_j^{-1/A_j} \right] \right) \nonumber \\
&= 1 - \mathrm{Prob}\left( \bigcup_{i\neq j}  \left[ s_i^{-1/A_i}  > s_j^{-1/A_j} \right] \right) \nonumber \\
&= 1 - \mathrm{Prob}\left( \bigcup_{i\neq j}  \left[ s_i  < s_j^{A_i/A_j} \right] \right).
\end{align}
Let $p_\mathrm{uni}(s)$ be a probability density function for the uniform random variable $U(0,1)$. Then,
\begin{align}
\mathrm{Prob}\left(  \left[ s_i  < s_j^{A_i/A_j} \right] \right) 
&= \int_0^1 ds_j \int_0^{s_j^{A_i/A_j}} ds_i \, p_\mathrm{uni}(s_i) p_\mathrm{uni}(s_j) \nonumber \\
&= \int_0^1 ds_j \, s_j^{A_i/A_j} \nonumber \\
&= \frac{A_j}{A_i + A_j} \nonumber \\
&= A_j B_i^{(j)},
\end{align}
where $B_i^{(j)} \equiv 1/(A_i + A_j)$. Similarly,
\begin{align}
&\mathrm{Prob}\left( \left[ s_{i_1}  < s_j^{A_{i_1}/A_j} \right] \bigcap \left[ s_{i_2}  < s_j^{A_{i_2}/A_j} \right] \right) \nonumber \\
&= \int_0^1 ds_j \, s_j^{A_{i_1}/A_j} s_j^{A_{i_2}/A_j} \nonumber \\
&= \frac{A_j}{A_{i_1} + A_{i_2} + A_j} \nonumber \\
&= A_j B_{i_1,i_2}^{(j)},
\end{align}
where $B_{i_1,i_2}^{(j)} \equiv 1/(A_{i_1} + A_{i_2} + A_j)$. Repeating the same procedures, we have 
\begin{align}
P_j &= 1 - \left( \sum_{i\neq j} A_j B_i^{(j)} - \sum_{i_1 < i_2, i_1 \neq j, i_2 \neq j} A_j B_{i_1,i_2}^{(j)} + \cdots \right) \nonumber \\
&= 1 - A_j \left( \sum_{i\neq j} B_i^{(j)} - \sum_{i_1 < i_2, i_1 \neq j, i_2 \neq j} B_{i_1,i_2}^{(j)} + \cdots \right) \nonumber \\
&= 1 - A_j B^{(j)},
\label{eq:interpretation_result}
\end{align}
where $B^{(j)} = \sum_{i\neq j} B_i^{(j)} - \sum_{i_1 < i_2, i_1 \neq j, i_2 \neq j} B_{i_1,i_2}^{(j)} + \cdots$.

The above discussion clarifies that the rule in Eq.~\eqref{select_rule} tends to select a spin that is difficult to flip. This is because the second term on the right-hand side of Eq.~\eqref{eq:interpretation_result} is proportional to $A_j$, and the index $j$ with a smaller $A_j$ is likely to be chosen. This fact is consistent with our intuition of the selection rule to escape from the local minima. Although the rule does not select the smallest $A_j$ because of factor $B^{(j)}$, the tendency to choose a spin with a smaller $A_j$ does not vary significantly. Hence, we conclude that the selection rule in Eq.~\eqref{select_rule} has an explicit interpretation consistent with our intuition.

Of course, it could be possible to apply other selection rules. We want to emphasize that the selection rule in Eq.~\eqref{select_rule} is analogous to the rejection-free selection rule. As mentioned above, the rejection-free selection rule is preferable because of its ease of implementation on hardware. Hence, we can expect that it will be easy to implement the proposed rule in Eq.~\eqref{select_rule} on hardware in the future. This is the main reason for utilizing this rule.

\section{Numerical Experiment}

Our proposed algorithm ideally performs best on specialized hardware, and its implementation is simple. Furthermore, it has natural interpretations, as discussed in Sect.~III.D. If the algorithm is convincing that it works well, users can use it with confidence. Here, we report the results of our basic numerical experiments to demonstrate the performance of the proposed method. After verification with an artificially generated problem, we conducted numerical experiments using the 0/1 knapsack problem as a problem instance that can be applied to a variety of real-world problems. In the numerical experiments, we compare the results of the conventional replica exchange method and those of the proposed method.

\subsection{Artificially generated problem}
\subsubsection{Problem settings}

As a toy model, we consider an artificial generation of problems. Since the aim is to yield a demonstration, we deal with a $30$-spin problem given in the Ising formulation. 

The values of the symmetric two-body interactions $\{J_{ij}\}$ are set as follows:
\begin{itemize}
  \item[1.] Generate a uniform random number $t$ for $t \in(-3,+3)$.
  \item[2.] Set $J_{ij} = t$ and $J_{ji} = t$
\end{itemize}
The values of the external magnetic field $h_i$ are determined as follows:
\begin{itemize}
  \item[1.] Generate a uniform random number $u$ for $u\in(-1,+1)$.
  \item[2.] If the sign of $u$ is positive, let $h_i = 2 + u$; if it is negative, let $h_i = -2 + u$.
\end{itemize}

We confirmed that the problems randomly generated via the above procedure have various energy landscapes: some are easy to find the global minimum, and some have many local minima. We consider only a small number of spins because our main target in this work is simply to check the proposed algorithm. Hence, it is easy to generate various problems and examine them in advance. Here, we choose one problem of moderate difficulty.

\subsubsection{Parameter setting}

In the replica exchange method, we use $M=5$ replicas. The temperature $T_m$ of the $m$-th replica is set as
\begin{align}
T_{m} = T_{\mathrm{min}} + T \left(\frac{m}{M}\right)^2,
\label{T_m}
\end{align}
where $T_{\mathrm{min}} = 10^{-3}$ and $T = 1$.

We perform $1000$ spin-flip trials of the Metropolis method in parallel for each replica. As for replica exchanges, two adjacent replicas $m, m+1$ are randomly selected once every $30$ flip trials. We exchange the two selected replicas with probability $P_{\mathrm{exchange}}^{(m,m+1)}$ in Eq.~\eqref{p_exchange}, and the exchange is rejected with probability $1 - P_{\mathrm{exchange}}^{(m,m+1)}$.

\subsubsection{Numerical results}

\begin{figure}
  \centering
  \includegraphics[keepaspectratio,width=85mm]{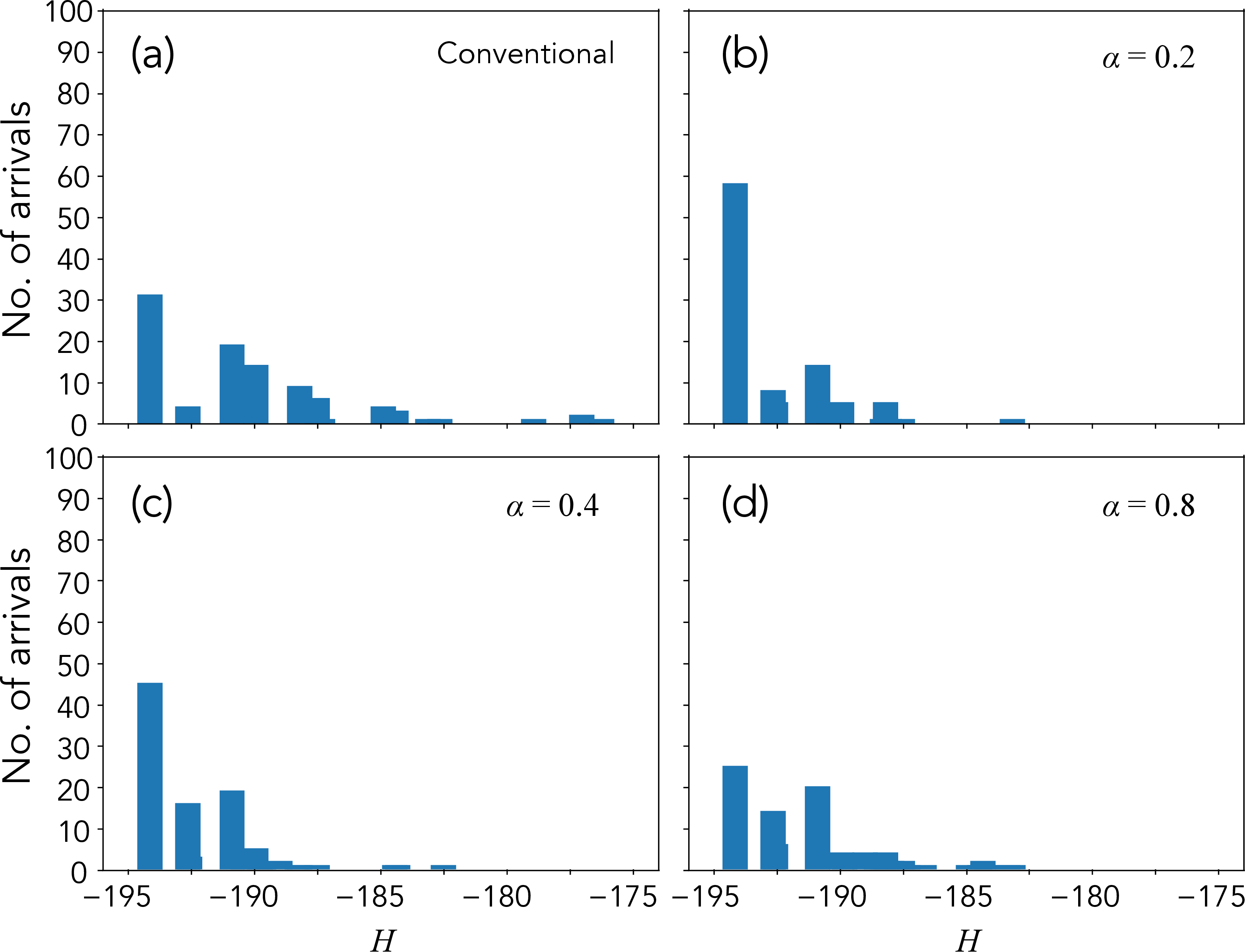}
\caption{Numerical results for optimization. We solved the same problem $100$ times. The horizontal axis of each figure represents the minimum energy in a trial, and the vertical axis represents the histogram. (a) shows the results of the conventional replica exchange method. (b) corresponds to the proposed method with the control parameter $\alpha = 0.2$. (c) and (d) correspond to cases with $\alpha = 0.4$ and $\alpha = 0.8$, respectively.}
  \label{fig:result}
\end{figure}

We solved the same optimization problem $100$ times using the proposed and conventional replica exchange methods. As explained in Sect.~III.C, trapping in a local minimum is judged to have occurred if there are $20$ successive rejections of the Metropolis rule. The proposed method has a control parameter $\alpha$ (see Algorithm~\ref{alg1}) that determines how often the forced moves are applied: namely, a large $\alpha$ means a large number of forced moves. In these experiments, we examined cases with $\alpha = 0.2, 0.4$, and $0.8$. As discussed in Sect.~III, the random re-initialization does not improve the optimization result. Hence, we expect that a too-large $\alpha$ may not be suitable because the effect is similar to that of random re-initialization.

Figure~\ref{fig:result} shows the optimization results, where the horizontal axis of each figure represents the minimum energy in a trial, and the vertical axis represents the histogram. It is easy to see that the proposed method with $\alpha = 0.2$ in Fig.~\ref{fig:result}(b) can find the lowest energy state many more times than the conventional method in (a). This result makes sense because the proposed algorithm enhances the escape from the local minima. In Fig.~\ref{fig:result}(b), (c), and (d), it seems that the effect of the escape decreases when $\alpha$ is relatively large. As discussed above, this is because the improvement due to random re-initialization is low. Since the results in Fig.~\ref{fig:result}(b), (c), and (d) are consistent with this finding, we can conclude that the proposed method behaved reasonably well.

We present the results for only one problem here, but we repeated the same experiments with different values of the two-body interactions and external magnetic fields and found that the results showed a similar tendency. Of course, the remaining task is to select an appropriate $\alpha$, which could be time-consuming, but several numerical experiments have indicated that a moderately wide range of $\alpha$s yields improvements. 

\subsection{Knapsack problem}

\subsubsection{Problem settings}

The knapsack problem is one of the famous combinatorial optimization problems in realistic situations. We here employ the instances for the 0/1 Knapsack problem by Ortega \cite{Ortega_online}. The problem statement is simple as follows:
\begin{itemize}
\item There are $N_\textrm{item}$ items. Item $i$ has its own value $v_i$ and weight $w_i$.
\item There is one knapsack. We can put items up to weight $W$ in the knapsack.
\item We want to make the total value of items in the knapsack as large as possible within the weight limit.
\end{itemize}
We use $N_\textrm{item}$ binary variables $\bm{z} = (z_1, \dots, z_{N_\mathrm{item}})$; item $i$ is packed in the knapsack ($z_i = 1$) or not ($z_i = 0$). Then, the problem becomes as follows:
\begin{itemize}
\item[] maximize $\displaystyle  \sum_{i=1}^{N_\mathrm{item}} v_i z_i $
\item[] subject to $\displaystyle  \sum_{i=1}^{N_\mathrm{item}} w_i z_i \leq W$.
\end{itemize}
Hence, we here employ the following energy function:
\begin{align}
E\left(\bm{z}, \bm{z}^\mathrm{s} \right)
&= - \sum_{i=1}^{N_\mathrm{item}} v_i z_i
+ \lambda \left( 
\sum_{i=1}^{N_\mathrm{item}} w_i x_i + \sum_{j=0}^{N_{\mathrm{W}}} 2^j z_j^{\mathrm{s}} - W
\right)^2,
\label{eq_knapsack_energy}
\end{align}
where $N_{\mathrm{W}} = \lfloor \log_2 W \rfloor$ and $\lambda$ is the penalty parameter for the weight constraint. The binary variables $\bm{z}^\textrm{s} = (z_1^\mathrm{s}, \dots, z_{N_{\mathrm{W}}}^\mathrm{s})$ are the slack variables for the weight constraint. (For the QUBO formulations with inequality constraints, for example, see Ref.~\cite{Lucas2014}.)

In the following numerical experiments, we use problem \verb|f2_l-d_kp_20_878| in the Ortega instances \cite{Ortega_online} as a demonstration; there are $N_\textrm{item}=20$ items and the weight limit is $W = 878$.

\subsubsection{Parameter settings}

With preliminary numerical experiments, we here use the following parameter settings.

The penalty parameter $\lambda$ in Eq.~\eqref{eq_knapsack_energy} is $\lambda = \max(v_i)+1$; in the problem setting of \verb|f2_l-d_kp_20_878|, $\max{v_i} = 91$ and then $\lambda = 92$. The control parameter in our proposed method, $\alpha$, is $\alpha = 0.4$. Other settings for the replica exchange method are the same as in Sect.~IV.A; we use $M = 5$ replicas. The temperature setting is also the same as Eq.~\eqref{T_m}.

\subsubsection{Numerical results}

\begin{figure}[b]
  \centering
\vspace{5mm}
  \includegraphics[keepaspectratio,width=85mm]{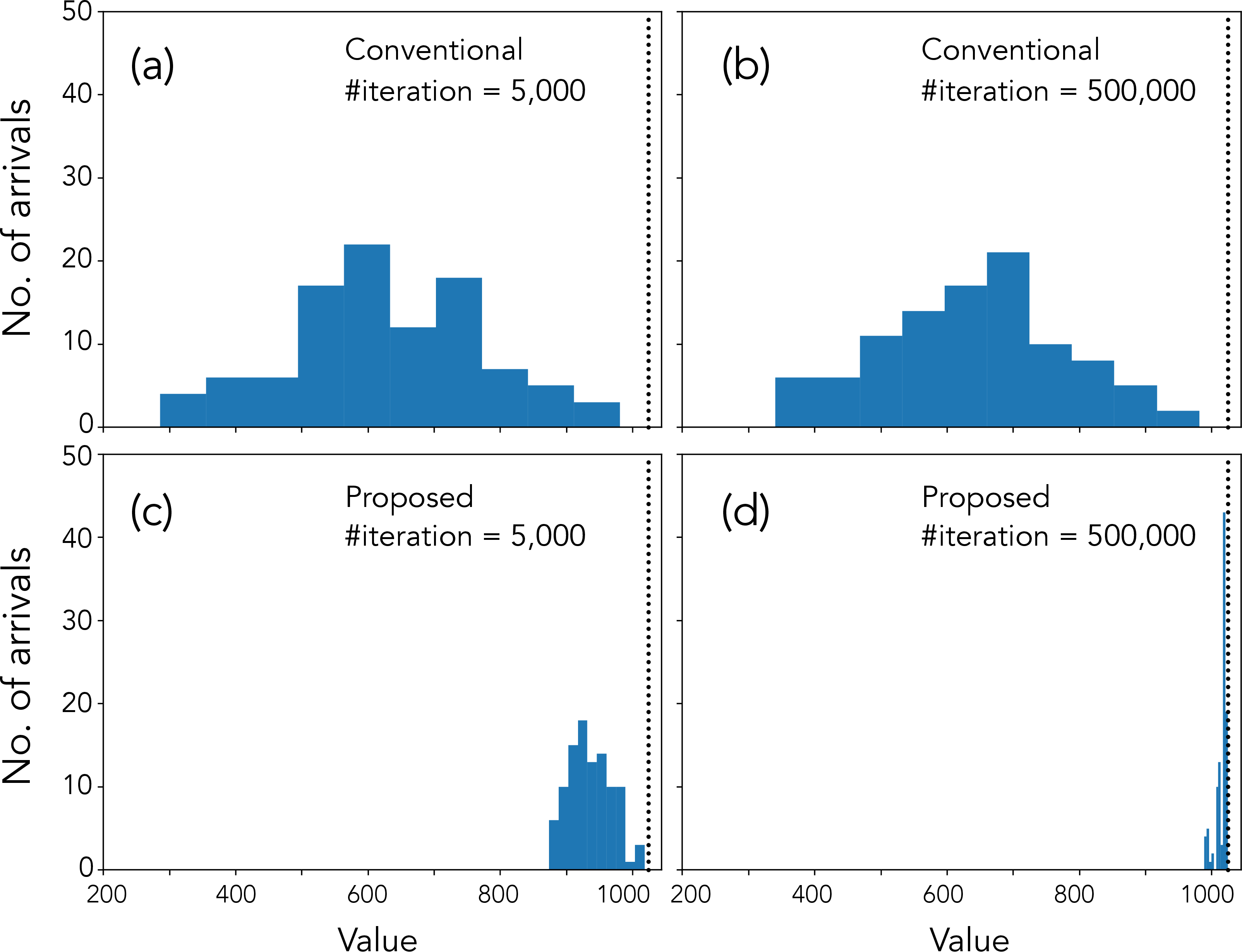}
\caption{
Numerical results for the knapsack problem. We solved the same problem $100$ times. The horizontal axis of each figure represents the total value of items, and the vertical axis represents the histogram. Note that the bin width for the histogram is adaptively changed. (a) and (b) show results by the conventional replica exchange method for the iteration number of $5,000$ and $500,000$ times, respectively. (c) and (d) are those with the proposed method. The dotted vertical line in each figure (value $= 1024$) means the maximum total value of the problem.}
  \label{fig:result_real}
\end{figure}

Figure~\ref{fig:result_real} shows the numerical results for the knapsack problem. Note that, different from Fig.~\ref{fig:result}, the horizontal axis means the total value of items, i.e. $\sum_i v_i z_i$. Hence, a higher value implies a better result. The maximum value of the total weight is depicted with the dotted vertical line in each figure. In addition, the bin width for the histogram is changed adaptively.

Although $5,000$ times iterations cannot find the optimal solutions with both the conventional and proposed methods, it is clear that the proposed method finds better results even with the small iteration numbers. After $500,000$ time iterations, the proposed method succeeds in finding the optimal solutions; in this experiment, the optimal solution was found $19$ times out of $100$ experiments. By contrast, the conventional method found only solutions far from optimal for 100 trials, even for the large iteration numbers. In addition, we checked that other choices of control parameter $\alpha$ yield similar results, and the improvement is observed for other knapsack problems by Ortega \cite{Ortega_online}.

As demonstrated here, the proposed method enhances the search for optimal solutions.

\vspace{5mm}

\section{Conclusion}

Local trapping in energy landscapes has long been problematic, and various methods for escaping from local minima have been proposed. In this paper, toward the goal of a future implementation using specialized hardware, we proposed a tractable method using an algorithm that is analogous to the rejection-free selection rule. As this algorithm is suitable for specialized hardware, the computational cost will be low. We also discussed our interpretation of the proposed algorithm, which naturally yields an escape from the local minima. For these reasons, the proposed algorithm is a strong candidate for future hardware implementations.

This paper represents a first step toward improving algorithms for efficiently solving combinatorial optimization problems. As a next step, it will be necessary to confirm whether the proposed method is effective even in large and complex combinatorial optimization problems. Of course, the relevant hardware implementations should also be investigated in the future.










\end{document}